\documentclass[pr1,longbibliography]{revtex4-1}
\usepackage{amsmath,amssymb,latexsym,epsfig,graphics,epsf}
\usepackage{hyperref}
\usepackage{graphics}
\usepackage{float}
\usepackage{footnote}
\usepackage[utf8]{inputenc}
\usepackage{adjustbox,amsmath,amsfonts}
\usepackage{blindtext}
\usepackage[T1]{fontenc}
\usepackage{lpic}
\usepackage{tabularx,booktabs}
\usepackage[dvipsnames]{xcolor}
\usepackage{ulem}
\newcommand{\be}{\begin{equation}}
	\newcommand{\ee}{\end{equation}}
\newcommand{\fig}[1]{Fig.~\ref{#1}}

\newcommand{\Fig}[1]{Figure~\ref{#1}}

\newcommand{\sect}[1]{Sec.~\ref{#1}}

\newcommand{\eq}[1]{Eq.~(\ref{#1})} 

\newcommand{\Sex}{{S}_{\rm ex}}
\newcommand{\Tc}{{T}_{\rm conf}}

\newcommand{\bF}{\mathbf F}

\newcommand{\bR}{\mathbf{R}}
\newcommand{\bRa}{{\bf R}_{\rm a}}
\newcommand{\bRb}{{\bf R}_{\rm b}}
\newcommand{\br}{\mathbf{r}}
\newcommand{\bn}{\mathbf{n}}

\newcommand{\LtoW}{{\rm LJ}\to{\rm WCA}}
\newcommand{\LtoI}{{\rm LJ}\to{\rm IPL}}
\newcommand{\YtoY}{{\rm YK}\to{\rm YK}}

\newcommand{\KtoL}{{\rm KA}\to{\rm mBLJ}}

\begin{document}
	\title{Interpolating between pair-potential systems}
	\date{\today}
	\author{Lorenzo Costigliola}\email{lorenzo.costigliola@gmail.com}
	\author{Andreas C. Martine, Claudia X. Romero, Jone E. Steinhoff,  Francisco M. F. A. S. da Fonseca, Maria B. T. Nielsen}
	\author{Jeppe C. Dyre}\email{dyre@ruc.dk}
	\affiliation{\textit{Glass and Time}, IMFUFA, Department of Science and Environment, Roskilde University, P.O. Box 260, DK-4000 Roskilde, Denmark}
	
	\begin{abstract}
		This paper studies liquid-model systems with almost identical constant-potential-energy hypersurfaces. We simulated continuous interpolations between such systems, specifically between the Lennard-Jones (LJ), Weeks-Chandler-Andersen (WCA), exponent 12 inverse-power-law (IPL), and Yukawa (YK) pair-potential systems. Structure and dynamics were monitored via the radial distribution function and the time-dependent mean-square displacement, respectively. In terms of the interpolation parameter, $0\le \lambda\le 1$, we argue that two systems have very similar constant-potential-energy hypersurfaces if the potential energies of configurations rarely cross when plotted as functions of $\lambda$. Such absence of ``level crossing'' applies to a very good approximation for the LJ to WCA transformation, and it also applies to a quite good approximation for the LJ to IPL and the YK to YK transformations (the latter varies the screening length). In all cases, structure and dynamics are shown to be almost invariant as functions of $\lambda$. The density is kept constant when $\lambda$ is varied. Temperature must generally be adjusted with $\lambda$, which is done by an iterative ``reduced-force-matching'' method with no free parameters. We also apply the interpolation strategy to two versions of the Kob-Andersen (KA) binary LJ system and show that a recently introduced shifted-force-cutoff version of this system has constant-potential-energy hypersurfaces, which are almost identical to those of the original KA system. This result rationalizes the previously established fact that the two KA versions have virtually identical physics.
	\end{abstract}
	\maketitle
	
	%\vspace{1cm}
	
	\section{Introduction}\label{sec:intro}
	
	The simplest model liquids are described by Newtonian dynamics and involve only pair-potential interactions \cite{bar76,han13}. Standard examples include the Lennard-Jones (LJ), inverse-power-law (IPL), and Yukawa (YK) pair-potential systems. It is always of interest when different systems have the same -- or almost the same -- structure and dynamics \cite{ros99,you03,you05,hey07,ram11,sch11,han13,lop13,dyr16}. A classical example of this is the observation that the radial distribution function of the LJ system is very close to that of a  hard-sphere system, an intriguing fact that was noted in 1968 in some of the earliest scientific computer simulations \cite{ver68} and which inspired the advent of successful perturbation theories based on the hard-sphere system as the zeroth-order approximation \cite{wca,bar76,han13,dym85,sar99,bom08,zho09,dub14}. 
	
	The term ``quasiuniversality'' describes the fact that many pair-potential systems have very similar structure and dynamics \cite{dyr16}. This is usually rationalized by reference to the hard-sphere system: whenever two systems at two thermodynamic state points are described well by hard-sphere systems of same packing fraction, the two systems have very similar physics. Incidentally, this reasoning was the background of Rosenfeld's ``excess-entropy scaling'' recipe from 1977 \cite{ros77}, which utilized the fact that the excess entropy is in a one-to-one correspondence with the packing fraction of the hard-sphere system mimicking the system in question. A recent alternative justification of quasiuniversality approximates the relevant pair potential as a sum of EXP (exponential repulsive) pair potentials \cite{bac14a}; by reference to isomorph theory \cite{IV,dyr14,dyr18a} such a sum leads to approximately the same structure and dynamics as that of a single EXP pair-potential system \cite{EXPII,dyr16}.
	
	In this paper, which builds on ideas of Ref. \onlinecite{dyr13}, we take an a different approach to predicting when one can expect two model liquids to have similar physics, an alternative that does not introduce a reference system. Our main proposition is that this is the case whenever the two systems have almost the same constant-potential-energy hypersurface, $\Omega$ \cite{dyr13}. If this applies, geodesic motion on the two $\Omega$s will be very similar. Since geodesic motion defines the so-called $NVU$ dynamics that leads to the same physics as standard Newtonian dynamics \cite{NVU_I,NVU_II}, the two systems must then have very similar structure and dynamics.
	
	We study below continuous interpolations between well-known pair-potential systems implemented by introducing a parameter $\lambda$ that varies from zero to unity. The investigation is divided into two parts. The first part shows results from computer simulations interpolating between different single-particle systems (\sect{sec:interpol}); a second part studies an interpolation between two binary systems (\sect{sec:KtoL}). In all cases we find that the constant-potential-energy hypersurface is almost independent of $\lambda$ and that structure and dynamics, as predicted \cite{dyr13}, are almost invariant with changing $\lambda$. In particular, this validates the $NVU$-based approach to the quasiuniversality of simple liquids \cite{dyr13}. The method used below is easily generalized to systems characterized by the non-spherically-symmetric interactions required to model molecular liquids.

	\section{Determining when different systems have similar physics}\label{sec:crit}
	
	Consider a system of $N$ point particles described by classical mechanics. The system's potential energy is denoted by $U(\bR)$ in which $\bR=(\br_1,...,\br_N)$ where $\br_i$ is the position vector of particle $i$. At a thermodynamic state point of temperature $T$, volume $V$, and (number) density $\rho=N/V$, the average potential energy is denoted by $\langle U\rangle$. The $3N-1$ dimensional \textit{constant-potential-energy hypersurface} $\Omega$ is defined by
	
	\be\label{Omega}
	\Omega
	\,=\,\{\bR\in \mathbb{R}^{3N}|U(\bR)=\langle U\rangle\}\,.
	\ee
	We regard the collective position vector $\bR$ as a point on a hypersurface of this kind. Usually one assumes periodic boundary conditions, in which case $\mathbb{R}^{3N}$ in \eq{Omega} is replaced by a $3N$-dimensional torus (periodic boundary conditions are used throughout this paper). For simplicity, the discussion is limited to interpolations between systems of same density, but generalization to systems of different densities is straightforward.
	
	$NVU$ dynamics is defined by replacing Newton's equation of motion by geodesic motion on $\Omega$. This can be shown to not change the structure and dynamics of the system in question \cite{NVU_I,NVU_II}. For instance, the time-autocorrelation functions of $NVU$ dynamics are identical to those of standard $NVE$ or $NVT$ dynamics (with obvious exceptions like the potential-energy time-autocorrelation function). As a consequence, if two systems have the same $\Omega$ -- possibly at state points with different temperatures -- then they will have the same physics (except for a possible scaling of time). Likewise, if they have \textit{approximately} same $\Omega$, the physics will be \textit{approximately} the same. By ``same physics'' we here mean same structure and dynamics while, e.g., thermodynamic properties like the average potential energy or the Helmholtz and Gibbs free energies are not expected to be identical (or even close). Note, however, that since the excess entropy is determined by the ``volume'' (area) of $\Omega$ -- specifically its logarithm  \cite{dyr13} -- this particular thermodynamic quantity \textit{is} invariant whenever $\Omega$ is.
	
	The mathematical criterion for two systems, ``$0$'' and ``$1$'', to have identical constant-potential-energy hypersurfaces is that for all configurations $\bRa$ and $\bRb$ 
	
	\be\label{eq:crit_eq}
	U_0(\bRa)=U_0(\bRb)\implies U_1(\bRa)=U_1(\bRb)\,.
	\ee
	The arrow $\implies$ here represents a logical implication in the mathematical understanding of this term. By symmetry, $\implies$ may be replaced by $\iff$. In practice, \eq{eq:crit_eq} is never rigorously obeyed for any two different systems. For this reason, it is more practical to be able to investigate whether two systems have \textit{approximately} the same $\Omega$, which is done by replacing the equality sign in \eq{eq:crit_eq} by $\cong$. This requires quantifying to which degree $\cong$ applies, however. An operational approach to check for approximately the same $\Omega$ is to study to which degree the following implication applies
	
	\be\label{eq:crit_eq2}
	U_0(\bRa)<U_0(\bRb)\implies U_1(\bRa)<U_1(\bRb)\,.
	\ee
	By going through the three different possibilities $U_0(\bRa)<U_0(\bRb)$, $U_0(\bRa)=U_0(\bRb)$, and $U_0(\bRa)>U_0(\bRb)$, it is straightforward to show that \eq{eq:crit_eq} and \eq{eq:crit_eq2} are mathematically equivalent. When it comes to numerical implementation, however, it makes better sense to check \eq{eq:crit_eq2} than \eq{eq:crit_eq} for being approximately obeyed \cite{lan25}. 
	
	What is the requirement on $U(\bR)$ in \eq{eq:crit_eq} and \eq{eq:crit_eq2}? This is simply that of thermodynamic consistency, i.e., that the system in question has a well-defined bulk thermodynamic limit and does not collapse with time. Exact mathematical criteria ensuring this were formulated long ago by Fisher and Ruelle \cite{fis66,hey07a}. In particular, they showed that if the spatial integral of the pair potential is negative, the system is unstable. This implies that if $U_0(\bR)$ is stable, a system with potential-energy function $-U_0(\bR)$ is unstable. Thus one cannot use in \eq{eq:crit_eq2} $U_1(\bR)=-U_0(\bR)$, which would violate the Fisher-Ruelle stability criterion (while obeying \eq{eq:crit_eq} by having the same constant-potential-energy hypersurface).
	
	Below we investigate how well \eq{eq:crit_eq2} applies by selecting a number of independent equilibrium configurations of system $0$ and plotting how their potential energies change when the system is gradually transformed into system $1$. If the curves generated in this way rarely cross, then \eq{eq:crit_eq2} applies to a good approximation and the two $\Omega$s must be almost identical. This approach is inspired by a similar procedure in isomorph theory \cite{sch14}, in which case it is the density of a given system that is changed, however, not the system itself. 
	
	As mentioned, we keep the density constant when comparing two systems. It is important to note that two systems may have the same $\Omega$ at state points of same density but \textit{different} temperatures. This is because the equilibrium temperature corresponding to a given $\Omega$ is neither determined by the value of the potential energy on $\Omega$ nor by the excess entropy $\Sex$ (which quantifies the volume of $\Omega$). Thus the temperature is given by the thermodynamic relation $T=(\partial U/\partial\Sex)_\rho$ \cite{IV,dyr13,dyr18a}, which can differ even for two systems with same $\Omega$. Note that this issue is only encountered because we perform $NVT$ simulations; the temperature does not need to be known if one instead uses $NVU$ dynamics \cite{NVU_I}.
	
	In order to check by $NVT$ simulations whether two systems have the same -- or very similar -- physics, we need to determine the equilibrium temperature at a state point with constant-potential-energy hypersurface $\Omega$. One way of doing this is by means of the so-called configurational temperature, $\Tc$, which in thermal equilibrium is identical to the temperature $T$ and is defined \cite{LLstat_3314,rug97,pow05} by
	
	\be\label{eq:Tc}
	k_B\Tc
	\,\equiv\,\frac{\langle(\nabla U)^2\rangle}{\langle\nabla^2U\rangle}\,.
	\ee
	Here the sharp brackets denote canonical averages, which in the thermodynamic limit $N\to\infty$ may be replaced by configuration-space microcanonical averages, i.e., averages over $\Omega$. In this limit, the relative fluctuations of $(\nabla U)^2$ and $\nabla^2 U$ go to zero, so for a large system one can estimate $\Tc$ reliably from a single equilibrium configuration $\bR\in\Omega$:
	
	\be\label{eq:Tc2}
	k_B T
	\,=\,k_B\Tc
	\,\cong\,k_B\Tc(\bR)
	\,\equiv\,\frac{(\nabla U(\bR))^2}{\nabla^2U(\bR)}\,.
	\ee
	The relative error in this estimate decreases with system size as $1/\sqrt{N}$. 
	
	An alternative to using \eq{eq:Tc2} for calculating the equilibrium temperature at a given state point is to employ so-called force matching \cite{sch22}. This approach has the advantage that it does not involve the Laplacian of $U$ but only forces, which are calculated anyway in a computer simulation. A disadvantage is that force matching in contrast to \eq{eq:Tc2} needs a reference system with same (or very similar) $\Omega$ for which the temperature is known. If such a system is at hand, one can calculate the ratio of the temperature of the system in question to that of the reference system from the following argument referring to the thermodynamic limit (for details please see Refs. \onlinecite{dyr13} and \onlinecite{sch22}): At any point $\bR$ on $\Omega$, the normal vector $\bn(\bR)$ is given by $\bn(\bR)=\nabla U(\bR)/|\nabla U(\bR)|$. The curvature $\kappa(\bR)$, which expresses how fast the normal vector changes when moving on $\Omega$, is given \cite{dyr13} by $\kappa(\bR)\propto\nabla\cdot\bn(\bR)\cong\nabla^2U(\bR)/|\nabla U(\bR)|\cong |\bF(\bR)|/\Tc(\bR)$ in which $\bF(\bR)=-\nabla U(\bR)$ is the $3N$-dimensional force vector and $\cong$ signals that the relative deviation vanishes in the thermodynamic limit. The equilibrium temperature $T$ of the system in question relative to that of the reference system, $T_0$, is determined by the fact that since the two systems have the same $\Omega$, one has $\kappa=\kappa_0$ implying that
	
	\be\label{eq:red_force_mat}
	T
	\,=\,T_0\,\frac{|\bF(\bR)|}{|\bF_0(\bR)|}\,. 
	\ee
	We refer to \eq{eq:red_force_mat} as ``reduced-force matching''; the determination of $T$ according to this is carried out iteratively (Appendix). Except for a density factor that is the same in all cases of this paper, the quantity $\bF (\bR)/k_BT$ is the so-called reduced force that plays an important role in isomorph theory \cite{IV,dyr14,dyr18a}.

	\section{Pair potentials and simulation details}\label{sec:ppot}

	\begin{figure}[H]
		\centering
		\includegraphics[width=0.4\linewidth]{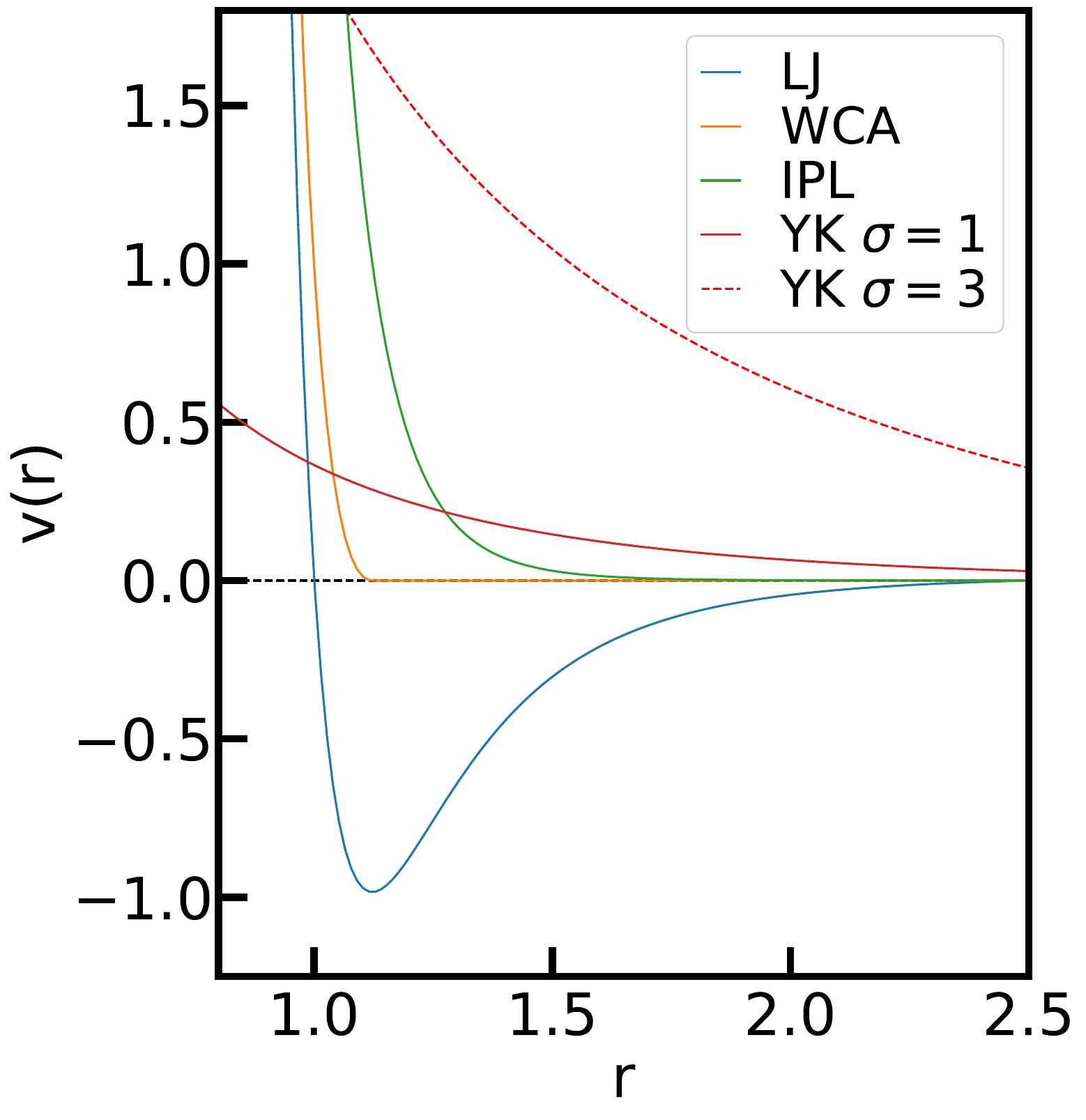}
		\caption{The pair potentials studied. Plotted as functions of the pair distance $r$, the figure shows the Lennard-Jones (LJ), Weeks-Chandler-Andersen (WCA), inverse power-law with exponent 12 (IPL), and Yukawa (YK) pair potentials. The Yukawa potential is shown for the two characteristic lengths, $\sigma=1$ and $\sigma=3$, between which we extrapolate (\fig{fig5}). }
		\label{fig1}
	\end{figure}
	
	We consider a systems of $N\gg 1$ particles of same mass, $m$. If $r_{ij}$ is the distance between particles $i$ and $j$, in terms of the  pair potential $v(r)$ the total potential energy $U$ is given by $U(\bR)=\sum_{i<j}v(r_{ij})$. This paper considers some of the most commonly studied pair potentials, the Lennard-Jones \cite{lj24}, Weeks-Chandler-Andersen \cite{wca}, inverse-power-law with exponent $12$ \cite{hoo71,bra06,hey07,hey08}, and Yukawa (YK) \cite{yuk35} potentials (\fig{fig1}). These are defined by

	\begin{figure}[H]
		\centering
		\includegraphics[width=\linewidth]{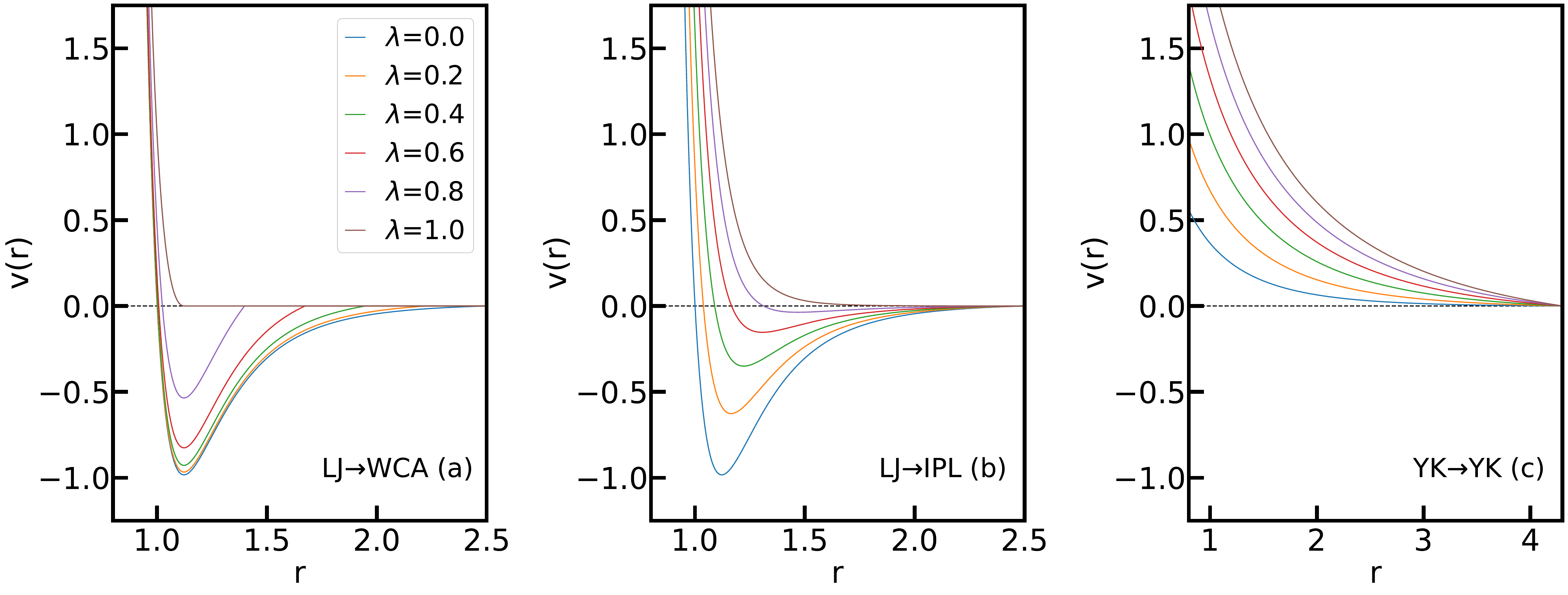}
		\caption{The three simulated pair-potential interpolations, each of which is shown for six values of $\lambda$ (legend of (a)). 
			(a) $\LtoW$, corresponding to cut-and-shifted LJ systems with a cutoff that decreases from 2.5 ($\lambda=0$, defining the standard LJ system) to $2^{1/6}$ ($\lambda=1$, defining the WCA system);
			(b) $\LtoI$;     
			(c) $\YtoY$.
			In all cases the potentials vary significantly so \textit{a priori} one might expect quite different physics.}
		\label{fig2}
	\end{figure}
	
	\begin{eqnarray}\label{eq:ppot}
		v_{\rm LJ}(r)&=& 4\varepsilon\,\left((r/\sigma)^{-12}-(r/\sigma)^{-6}\right)\nonumber\\
		v_{\rm WCA}(r)&=& 
		\begin{cases}v_{\rm LJ}(r)
			-v_{\rm LJ}(2^{1/6}\sigma)\,\,,\,\,r<2^{1/6}\sigma\nonumber\\
			0\,\,,\,\,r>2^{1/6}\sigma
		\end{cases}\\
		v_{\rm IPL}(r)&=& \varepsilon\,(r/\sigma)^{-12}\\
		v_{\rm YK}(r)&=& \frac{\varepsilon}{(r/\sigma)}\,e^{- r/\sigma}\nonumber\,.
	\end{eqnarray}
	In these expressions $\varepsilon$ determines the energy scale and $\sigma$ the length scale of the pair interaction. Unless otherwise stated we use the rationalized unit system in which $\varepsilon=1$, $\sigma=1$, and $k_B=1$. When evaluating how much structure and dynamics vary when interpolating between two pair potentials, the density $\rho$ is kept constant while temperature is adjusted as described in \sect{sec:crit} and the Appendix.
	
	A general method of interpolating between system $0$ of pair potential $v^{(0)}(r)$ and system $1$ of pair potential $v^{(1)}(r)$ uses the convex combination $v^{(\lambda)}(r)=(1-\lambda) v^{(0)}(r) + \lambda v^{(1)}$ ($0\leq \lambda\leq 1$), which generates a one-parameter family of pair potentials that clearly interpolates smoothly between systems $0$ and $1$. Below we take this approach in some cases, but generally employ the philosophy of implementing the most natural interpolation. In this connection, recall that in their 1971 paper \cite{wca}, Weeks, Chandler, and Andersen regarded the LJ pair potential as a sum of a purely repulsive (``WCA'') and an attractive term, which suggests that one way of interpolating from LJ to WCA is by gradually removing the attractive term. We instead utilize the fact that the WCA pair potential is defined as the LJ pair potential cut and shifted at the potential-energy minimum, and gradually reduce the cutoff from the standard $r_c=2.5$ to $r_c=2^{1/6}$ that identifies the LJ pair-potential minimum.
	
	All simulations were carried out using the GPU-optimized Molecular Dynamics code RUMD \cite{RUMD} with standard Nose-Hoover $NVT$ dynamics \cite{tildesley}. For the single-component systems $N=4096$ particles were simulated, for the binary systems $N=4000$, in both cases at typical liquid state points. Further details about the simulations are given in the Appendix.

	\section{Interpolating between single-particle systems}\label{sec:interpol}
	
	We transform LJ gradually into WCA, denoted by $\LtoW$, by reducing the standard shifted-potential cutoff from 2.5 to $2^{1/6}$, i.e.,
	
	\be\label{eq:LtoW}
	v_{\LtoW}^{(\lambda)}(r)
	\,=\,
	\begin{cases}
		v_{\rm LJ}(r)-v_{\rm LJ}(r_c(\lambda))\,\,&,\,\,r<r_c(\lambda)\equiv(1-\lambda)2.5+\lambda 2^{1/6}\\
		0\,\,&,\,\, r>r_c(\lambda)\,.
	\end{cases}
	\ee
	Clearly, $v_{\LtoW}^{(0)}(r)=v_{\rm LJ}(r)$ with a shifted-potential cutoff at 2.5 and $v_{\LtoW}^{(1)}(r)=v_{\rm WCA}(r)$.  
	
	The second interpolation is between LJ and IPL, which is carried out by gradually removing the attractive $r^{-6}$ term:
	
	\be\label{eq:LtoI}
	v_{\LtoI}^{(\lambda)}(r)
	\,=\,4\,\left(r^{-12}\,-\,(1-\lambda)\, r^{-6}\right)\,.
	\ee
	This expression has the required limits $v_{\LtoI}^{(0)}(r)=4\left(r^{-12}\,-\, r^{-6}\right)$ and $v_{\LtoI}^{(1)}(r)=4 r^{-12}$.
	
	The third case interpolates between two Yukawa pair potentials of different screening length. This is done by changing the parameter $\sigma$ in \eq{eq:ppot} from 1 to 3, i.e., by putting $\sigma=1+2\lambda$:
	
	\be\label{eq:YtoY}
	v_{\YtoY}^{(\lambda)}(r)
	\,=\,\frac{1+2\lambda}{r}\,e^{-r/(1+2\lambda)}\,.
	\ee
	\Fig{fig2} shows the pair potentials of these interpolations for six values of $\lambda$. Note that the pair potentials vary considerably with $\lambda$.

	\begin{figure}[H]
		\centering
		\includegraphics[width=\linewidth]{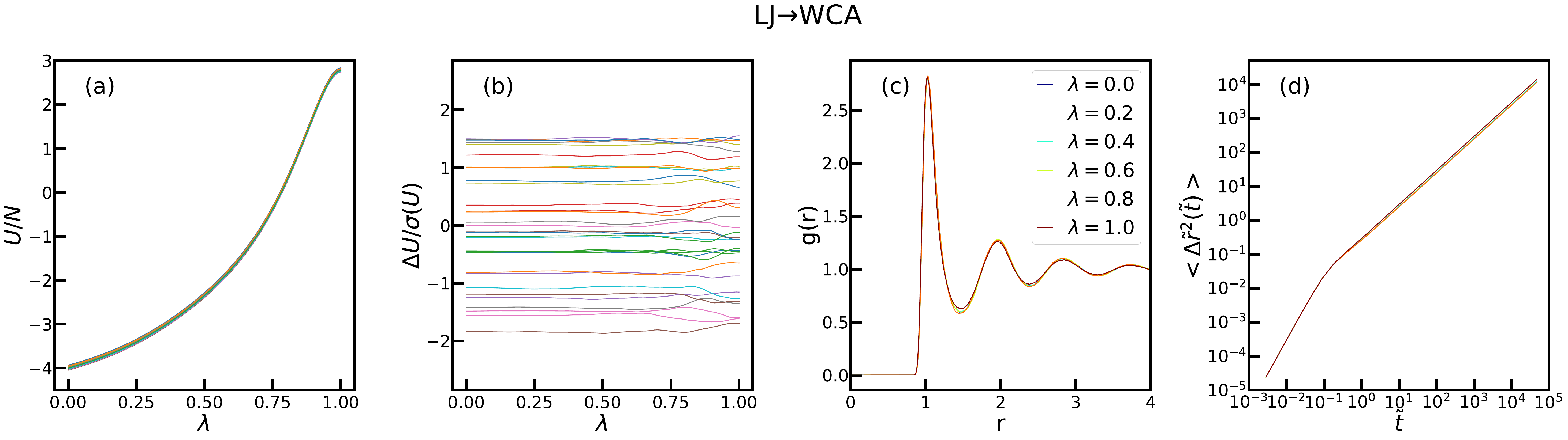}
		\caption{Results for the $\LtoW$ interpolation. The reference state point $(\rho, T)=(1.0,2.0)$ is where the $\lambda=0$ simulation was carried out, which forms the basis for the $\lambda$-scaling plots in (a) and (b).
			(a) Per-particle potential energy with changing $\lambda$ plotted for 3 equilibrium configurations.
			(b) Fluctuations of $U$ around the sample-average for 32 independent configurations, scaled with the standard deviation $\sigma(U)$. For $\lambda < 0.75$ there are virtually no ``level crossings'', and for $\lambda>0.75$ \eq{eq:crit_eq2} is still largely obeyed.
			(c) Radial distribution functions $g(r)$ for six $\lambda$ values. For each $\lambda$ the temperature was adjusted according to the reduced-force-matching criterion \eq{eq:red_force_mat}, leading only to minor temperature changes (see below).
			(d) Reduced-unit mean-square displacement as a function of time with changing $\lambda$.
		}\label{fig3}
	\end{figure}

	Consider first $\LtoW$ (\fig{fig3}). Panel (a) shows how the potential energy varies with $\lambda$ for 32 independent configurations generated in a standard equilibrium Nose-Hoover $NVT$ simulation of the LJ system (defining $\lambda=0$) at the reference thermodynamic state point $(\rho,T)=(1.0,2.0)$. No extra simulations were carried out in order to calculate how the potential energies of these configurations change with $\lambda$, which is determined entirely by the pair potential $v_{\LtoW}^{(\lambda)}(r)$ and the configuration in question. Not surprisingly, the potential energy increases when the shifted-potential cutoff is reduced; in fact it changes from negative to positive values. It is not possible to judge from panel (a) to which degree \eq{eq:crit_eq2} is obeyed, however, because the potential energies of the 32 configurations follow each other closely. To investigate the \eq{eq:crit_eq2} criterion we plot in panel (b) as a function of $\lambda$ the normalized potential-energy variation, $\Delta U/\sigma(U)$, in which $\Delta U=U-\langle U\rangle$ and $\sigma^2(U)\equiv\langle(\Delta U)^2\rangle=\langle U^2\rangle-\langle U\rangle^2$ (averaging over the 32 configurations). If these curves never intersect -- no ``level crossings'' -- then \eq{eq:crit_eq2} is rigorously obeyed. This is not rigorously the case, but it does apply to a very good approximation. Consequently, as argued above by reference to $NVU$ dynamics, one expects very similar structure and dynamics. We investigate this by carrying out $NVT$ simulations for $\lambda=0.0, 0.2, 0.4, 0.6, 0.8, 1.0$ (all at density $1.0$), with temperatures adjusted by reduced-force matching (\sect{sec:crit} and the Appendix). The resulting radial distribution function (RDF) and the reduced-unit mean-square displacement (MSD) as a function of time are shown in panels (c) and (d). The structure and dynamics of all systems are clearly very similar. This is not at all unexpected in view of the well-known fact that the WCA system provides an excellent approximation to the LJ system \cite{wca,han13}, which is further confirmed by the fact that the temperature adjustments all are below 5\% (in fact, not adjusting the temperature gives virtually as good results).
	
\begin{figure}%[H]
		\centering
		\includegraphics[width=0.6\linewidth]{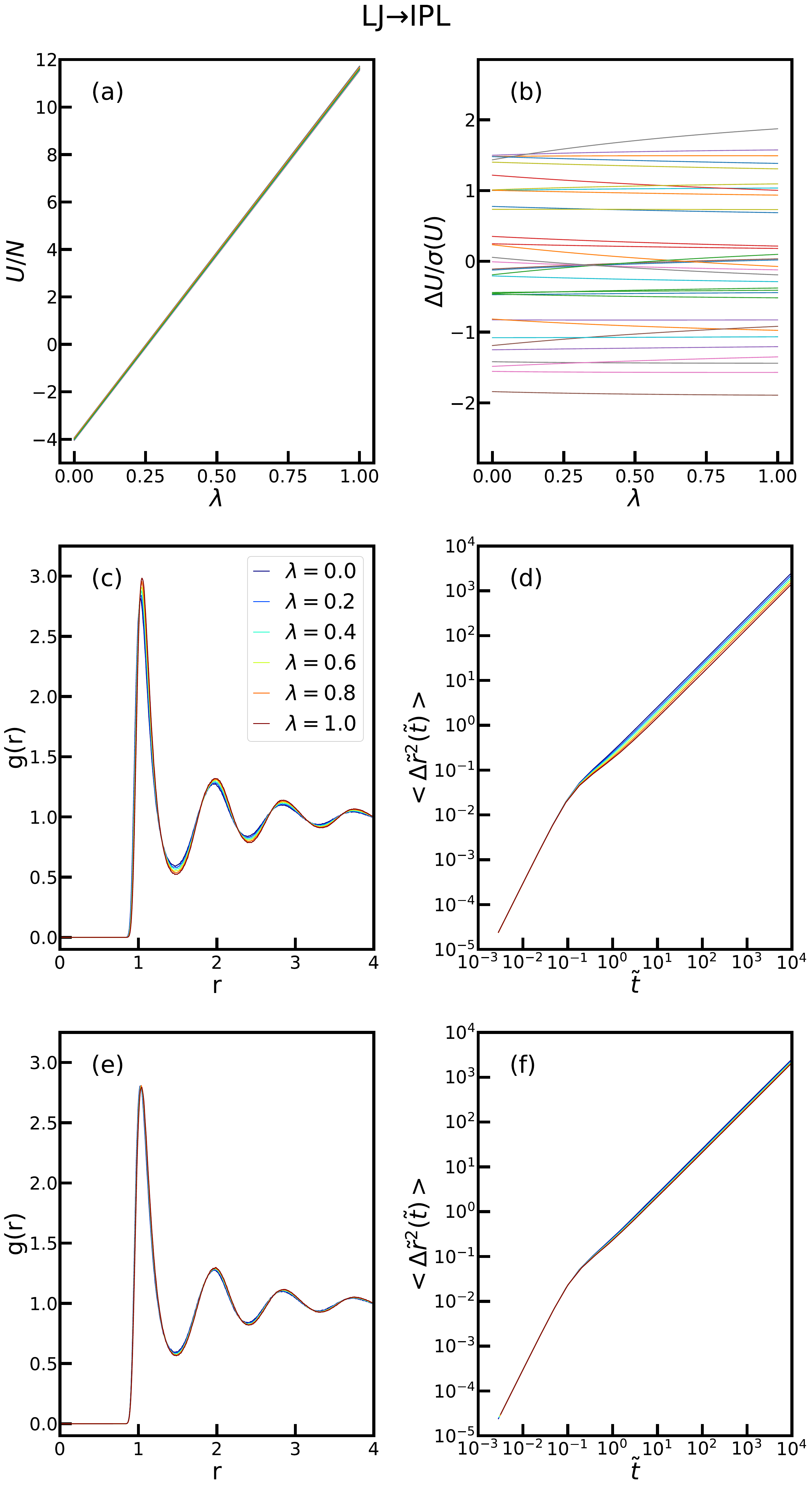}
		\caption{Results for $\LtoI$ with reference state point $(\rho, T)=(1.0,2.0)$. 
			(a) and (b) are analogous to the same panels in \fig{fig3}.
			(c) and (d) show the variation of structure and dynamics if no temperature adjustment is implemented, in which case the physics is not quite invariant. 
			(e) and (f) show structure and dynamics when the temperature is adjusted according to the reduced-force matching method (Appendix). Here both structure and dynamics are invariant to a notably higher degree.}
		\label{fig4}
	\end{figure}

	The $\LtoI$ interpolation is studied in \fig{fig4} at the same reference state point, $(\rho,T)=(1.0,2.0)$. Although the IPL and LJ pair potentials are quite different, (b) shows that \eq{eq:crit_eq2} is obeyed to a quite good approximation. Panels (c) and (d) show structure and dynamics at the temperature of the reference state point,  $T=2.0$, for varying $\lambda$. Panels (e) and (f) show analogous data with temperature adjustment, leading to almost perfect data collapse. We return to the magnitude of the temperature adjustment in \fig{fig6} below.

	\begin{figure}%[H]
		\centering
		\includegraphics[width=0.6\linewidth]{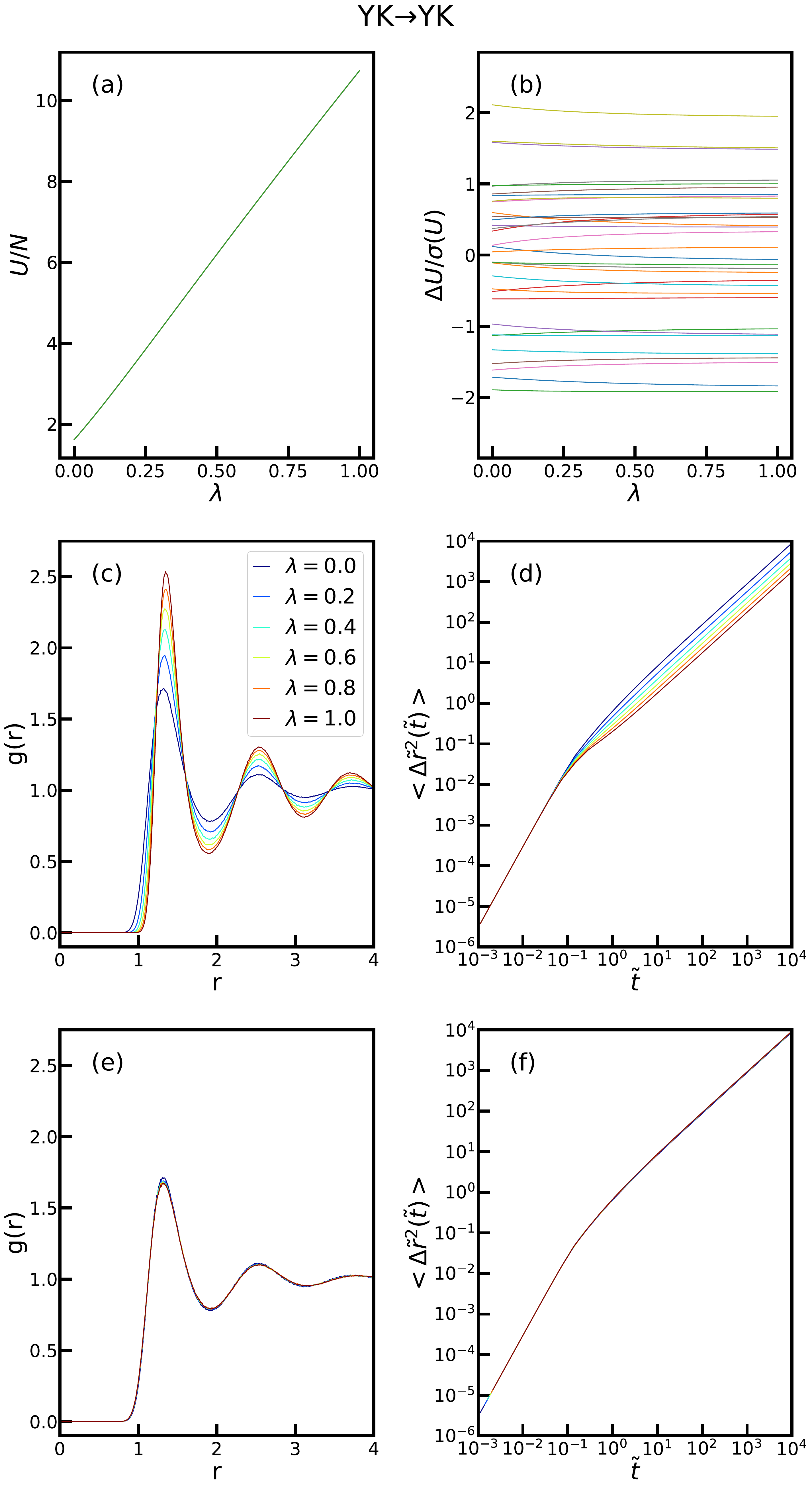}
		\caption{Results for $\YtoY$ with reference state point $(\rho, T)=(0.5,0.02)$. 
			(a) and (b) are analogous to the same panels in \fig{fig3}.
			(c) and (d) show the structure and dynamics if no temperature adjustment is carried out. The physics changes a lot with $\lambda$.
			(e) and (f) show the same with temperature adjusted, in which case structure and dynamics are invariant to a very good approximation.}
		\label{fig5}
	\end{figure}

	We finally turn to $\YtoY$, in which case the reference state point is $(\rho,T)=(0.5,0.02)$. As previously, we performed a $\lambda=0$ simulation to generate the 32 configurations used to calculate the potential energies as functions of $\lambda$ according to \eq{eq:YtoY}. \Fig{fig5}(a) shows that the potential energies of these configurations scale in a very similar way when $\lambda$ is increased. Again, this does not allow for checking how well \eq{eq:crit_eq2} applies. Panel (b) shows few level crossings. Indeed, structure and dynamics are almost independent of $\lambda$ for the temperature-adjusted simulations (panels (e) and (f)), which is in notable contrast to what happens if the temperature is not adjusted (panels (c) and (d)).

	\begin{figure}[H]
		\centering
		\includegraphics[width=\linewidth]{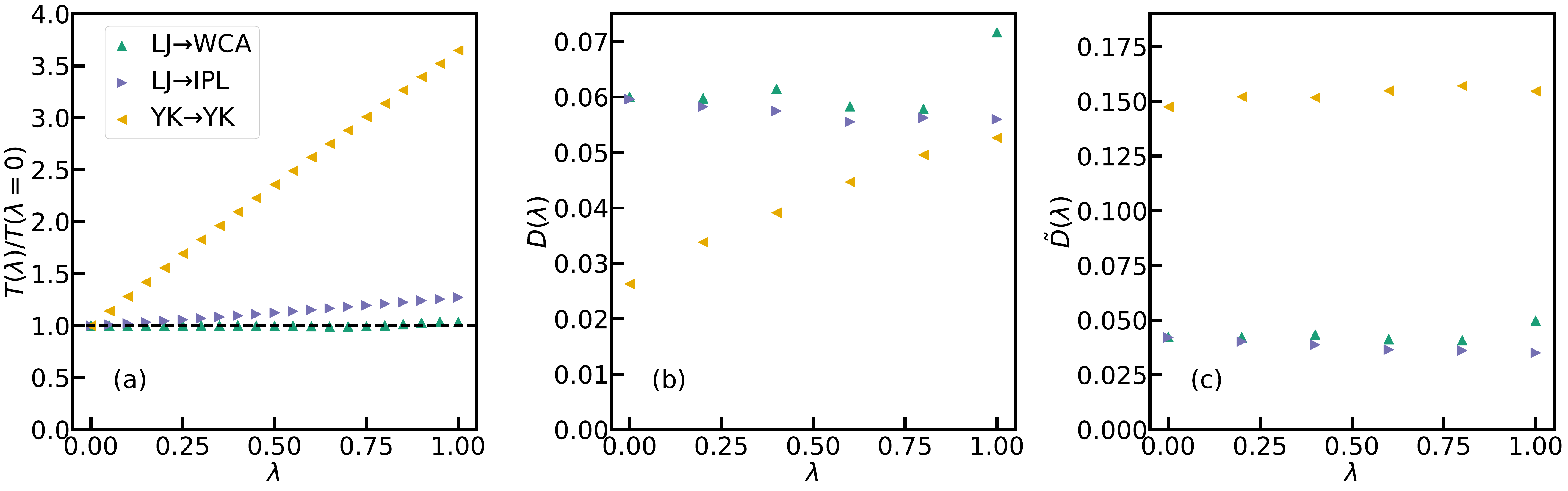}
		\caption{Temperature adjustment.
			(a) shows the temperatures determined by reduced-force matching (\eq{eq:red_force_mat}) as functions of $\lambda$; the black dashed horizontal line marks unity. In the $\LtoW$ case the effect is very small, which confirms the well-known fact that the LJ and WCA systems have very similar physics at a given state point \cite{wca}. In the $\LtoI$ case up to 25\% temperature adjustment is needed, and even larger adjustments are necessary for $\YtoY$.
			(b) shows the diffusion coefficients $D$ as a function of $\lambda$ obtained from the long-time MSD.
			(c) shows analogous data for the reduced diffusion coefficient $\tilde D$ with the $T(\lambda)$ temperature adjustment in $NVT$ simulations. The reduced diffusion coefficients are almost constant with changing $\lambda$.}
		\label{fig6}
	\end{figure}
	
	For the three interpolations \fig{fig6}(a) shows the reduced-force-matching temperatures relative to the reference temperature. \Fig{fig6}(b) shows the variation with $\lambda$ of the diffusion coefficient, $D$, and (c) shows how much the reduced diffusion coefficient defined by $\tilde{D}\equiv\rho^{1/3}\sqrt{m/k_BT}D=\rho^{1/3}T^{-1/2}D$ \cite{ros77,IV} varies with $\lambda$ (the last equality reflects the use of rationalized units). $\tilde{D}$ is almost constant. This confirms the invariance of the dynamics when temperature adjustment is implemented.

	\section{Interpolating between two versions of a binary LJ system}\label{sec:KtoL}
	
	In order to illustrate the generality of our method, this section studies a transformation involving the well-known Kob-Andersen (KA) binary LJ mixture \cite{kob95}. The KA system, which consists of 80\% A particles and 20\% B particles, is resistant against crystallization and therefore easily supercooled \cite{kob95,ped18}. For this reason the KA system has become a standard model for simulation studies of glass-forming liquids \cite{fra02,cro15,cos18a,ped18}. Writing the LJ pair potential between particles of type $\alpha$ and $\beta$ as $v_{\alpha \beta}(r)=4 \varepsilon_{\alpha \beta}((r/\sigma_{\alpha \beta})^{-12}-(r/\sigma_{\alpha \beta})^{-6})$, in which $\alpha$ is A or B and likewise for $\beta$, the parameters defining the KA system are \cite{kob95}: $\sigma_{AA}=1.0$, $\sigma_{AB}=\sigma_{BA}=0.8$, $\sigma_{BB}=0.88$, $\varepsilon_{AA}=1.0$, $\varepsilon_{AB}=\varepsilon_{BA}=1.5$, $\varepsilon_{BB}=0.5$.

	\begin{figure}[H]
		\centering
		\includegraphics[width=0.6\linewidth]{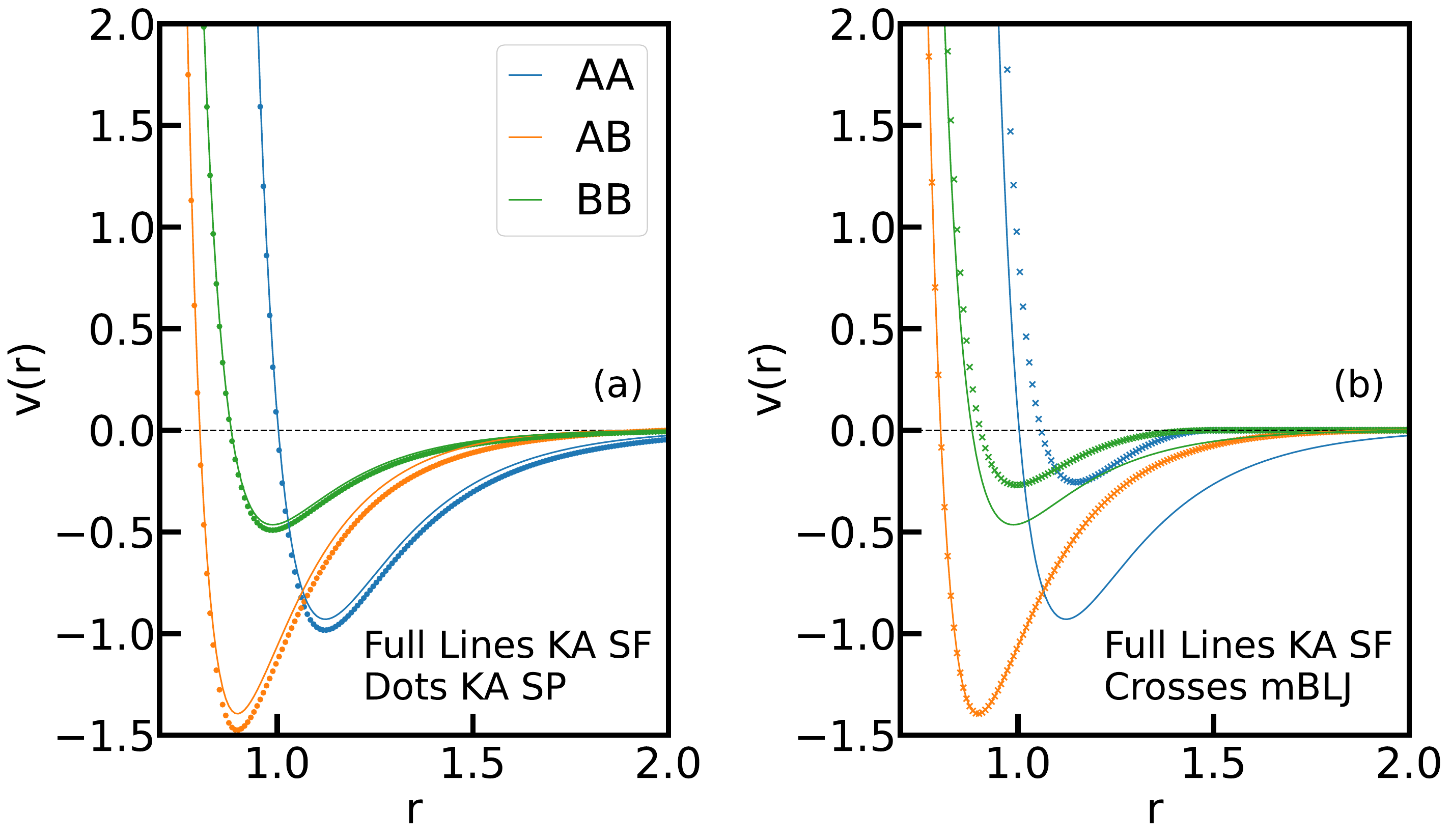}
		\caption{Binary LJ pair potentials.
			(a) shows the potentials of the standard, shifted-potential (SP) cutoff version of the KA model (dots) and the shifted-force (SF) version with the same cutoffs (full curves). The two cases have almost the same pair potentials, which is a consequence of the relatively large cutoff used.
			(b) compares the SF version of the KA system with its modified version (mBLJ) that has significantly smaller like-particle cutoffs.}
		\label{fig7}
	\end{figure}

	We interpolate between the standard KA model and a novel version of it, the so-called modified binary LJ (mBLJ) model that is at least 100 times less prone to crystallization \cite{sch20}. This is achieved by changing the like-particle pair-potential cutoffs to shifted-force cutoffs at significantly smaller distances. Recall that a shifted-force cutoff at $r_c$ replaces  the pair force $f(r)\equiv -v'(r)$ by $f(r)-f(r_c)$ below $r_c$ \cite{tildesley,tox11a}, a procedure that results in much better energy conservation than the standard shifted-potential cutoff because there is no pair-force discontinuity at $r_c$. In the present case, the primary effect of switching to short-range shifted-force cutoffs is that the like-particle attractive interactions become much weaker than in the standard KA model thus preventing crystallization of the KA system (which proceeds via phase separation and subsequent creation of a pure A particle face-centered cubic crystal \cite{ped18}). Specifically, the mBLJ model has like-particle shifted-force cutoffs at $r_c=1.5\sigma_{AA}$, while a shifted-force cutoff at $r_c=2.5\sigma_{AB}$ is used for the AB interaction. 
	
	The standard KA model involves cutoffs at $r_c=2.5\sigma_{\alpha\beta}$. One usually implements these as shifted-potential cutoffs, but for technical reasons we use shifted-force cutoffs also to represent the standard KA model. The KA and mBLJ pair potentials with different cutoffs are shown in \fig{fig7}. Panel (a) compares the shifted-potential and shifted-force versions of the standard KA system of $r_c=2.5\sigma_{\alpha\beta}$ cutoffs. We see that there is little difference, and as shown in the Appendix the dynamics of the two systems are indeed very similar. This justifies our use of the shifted-force KA version to represent the standard KA model. Panel (b) shows the shifted-force KA and mBLJ pair potentials. The like-particle attractions are much weaker in the mBLJ case than in the KA case, in particular for the A particles.

	\begin{figure}[H]
		\centering
		\includegraphics[width=\linewidth]{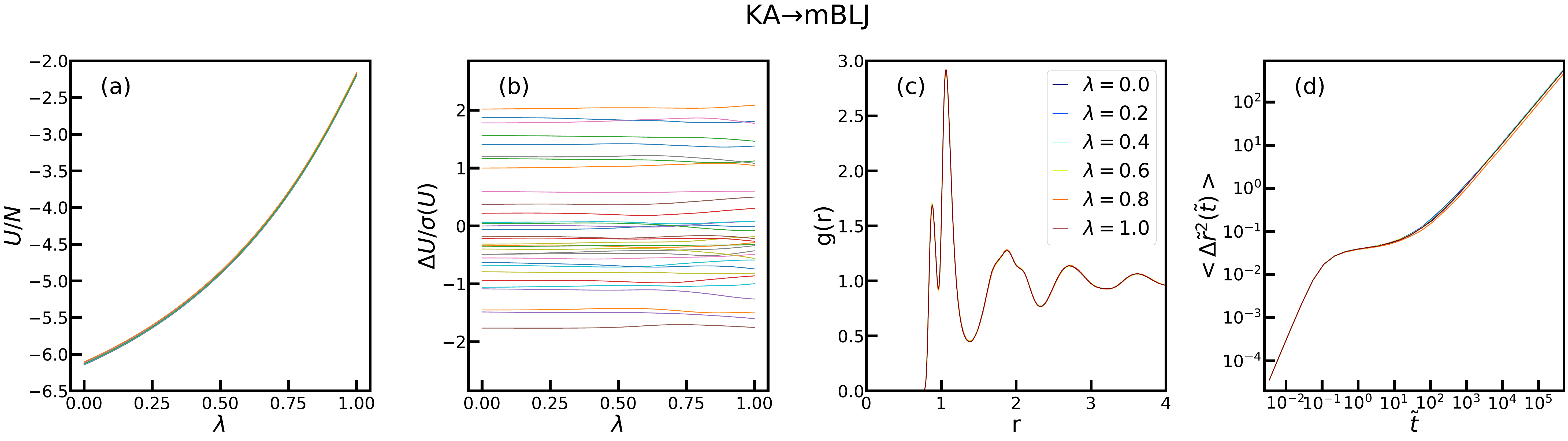}
		\caption{Results for the binary system interpolation $\KtoL$, plotted as in \fig{fig3}. Just as in the single-component cases, the invariance of $\Omega$ that is evident from (b) translates into invariance of the structure and the dynamics. 
			(c) and (d) refer to all-particle data with temperature adjustment, i.e., the A and B particles are not distinguished here.}
		\label{fig8}
	\end{figure}

	The interpolation from the standard KA model to the mBLJ model is carried out by using shifted-force cutoffs that are a convex combination of the old and new ones, i.e., for each of the three pair interactions the cutoff $r_c(\lambda)$ is written as $1-\lambda$ times the KA cutoff plus $\lambda$ times the mBLJ cutoff. For the like-particle interactions this amounts to $r_c(\lambda)=(2.5-\lambda)\sigma_{AA}$ for the A  particles and $r_c(\lambda)=2.5(1-\lambda)\sigma_{BB}+1.5\lambda\sigma_{AA}$ for the B particles; for the AB interaction the cutoff remains at $2.5\sigma_{AB}$. \Fig{fig8} shows the results of this transformation. The reference state point is $(\rho, T)=(1.2,0.48)$, which is in the less-viscous liquid phase compared to the lower temperatures of current numerical studies of glass-forming liquids. As previously, panel (a) shows the potential-energy variation as a function of $\lambda$ and (b) shows the relative variations. There are few level crossings. Consistent with this (c) and (d) show invariant all-particle structure and dynamics (temperature-adjusted data).

	\begin{figure}[H]
		\centering
		\includegraphics[width=0.8\linewidth]{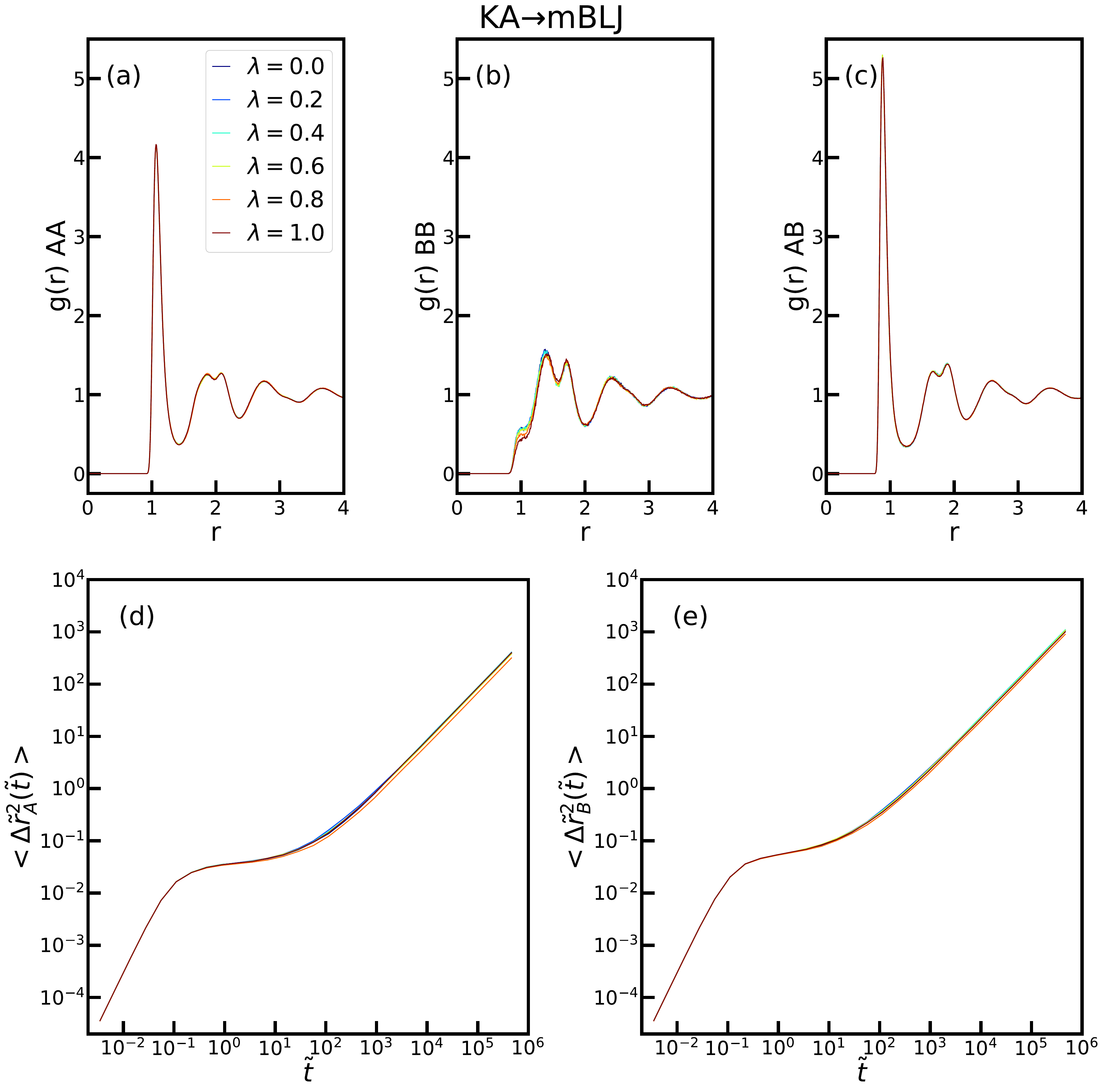}
		\caption{Closer analysis of the $\KtoL$ interpolation in which (a), (b), and (c) report RDF data while (d) and (e) report MSD data. Overall there is a good data collapse, with a minor deviation for $\lambda=0.8$ (compare \fig{fig10} below).}
		\label{fig9}
	\end{figure}

	Given that the KA system consists of two different particle types, one may ask whether the structure and dynamics of A and B particles are separately invariant. This is investigated in \fig{fig9}. Overall we see good invariance.

	\begin{figure}[H]
		\centering
		\includegraphics[width=\linewidth]{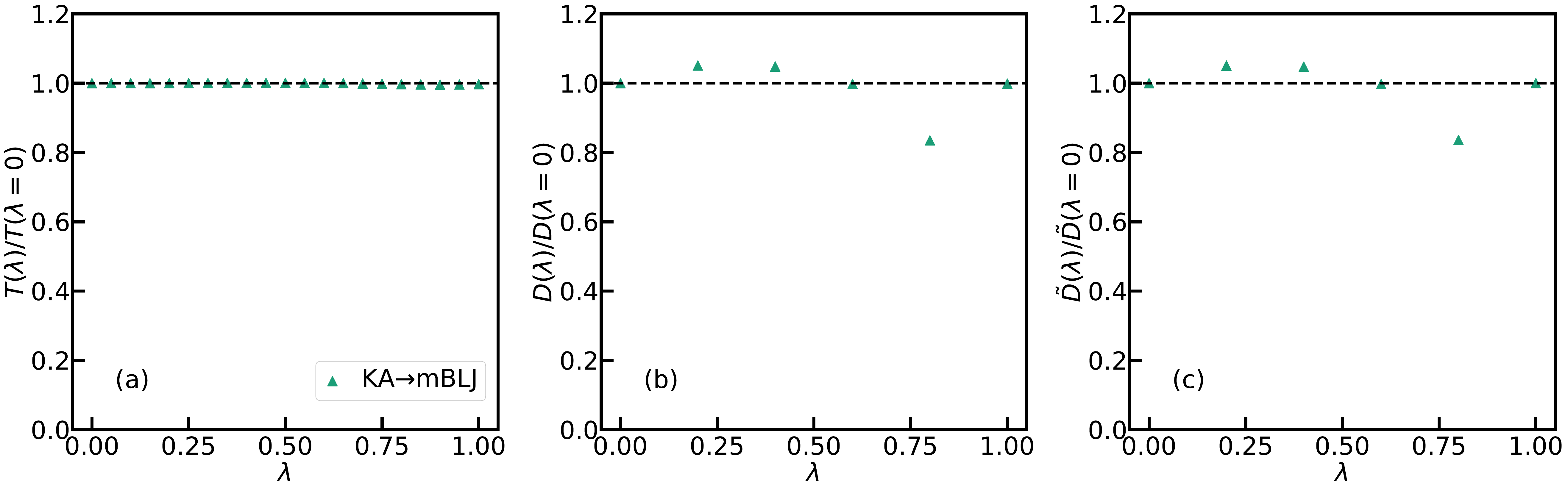}
		\caption{Effect of temperature adjustment in the $\KtoL$ case. 
			(a) shows that only very little adjustment is needed. 
			(b) and (c) show data for the all-particle diffusion coefficient and its reduced-unit version.
		}\label{fig10}
	\end{figure}

	We finally show in \fig{fig10}(a) that virtually no temperature adjustment is needed in the $\KtoL$ case. This important result is consistent with extensive simulations carried out at $T=0.37$ at which the system is strongly supercooled; here the KA and mBLJ systems were also found to have virtually the same dynamics \cite{sch20}. The relative variation of the diffusion coefficient and its reduced version are shown in (b) and (c), respectively. There is little variation although we note a drop at $\lambda=0.8$ for which we have no explanation. Table 1 reports the non-reduced diffusion coefficients for all interpolations.

	\begin{table}
		\begin{center}
			\caption{Diffusion coefficient $D$ as a function of $\lambda$ (left column) for the four system considered.}
			\begin{tabular}{ccccc}
				\toprule
				$\lambda$ & \multicolumn{4}{c}{$D$} \\
				\cmidrule(l){2-5}
				\phantom{$\lambda$} & ${\rm LJ}\to{\rm WCA}$ & ${\rm LJ}\to{\rm IPL}$ & ${\rm YK}\to{\rm YK}$ & ${\rm KA}\to{\rm mBLJ}$ \\
				\hline
				0.00 & 0.0600 & 0.0596 & 0.0263 & 0.000129 \\
				0.20 & 0.0598 & 0.0583 & 0.0338 & 0.000135 \\
				0.40 & 0.0615 & 0.0575 & 0.0391 & 0.000135 \\
				0.60 & 0.0583 & 0.0555 & 0.0447 & 0.000128 \\
				0.80 & 0.0578 & 0.0563 & 0.0496 & 0.000107 \\
				1.00 & 0.0717 & 0.0560 & 0.0526 & 0.000128 \\
				\bottomrule
			\end{tabular}
		\end{center}
	\end{table}

	\section{Discussion}
	
	\begin{figure}[H]
		\centering
		\includegraphics[width=0.6\linewidth]{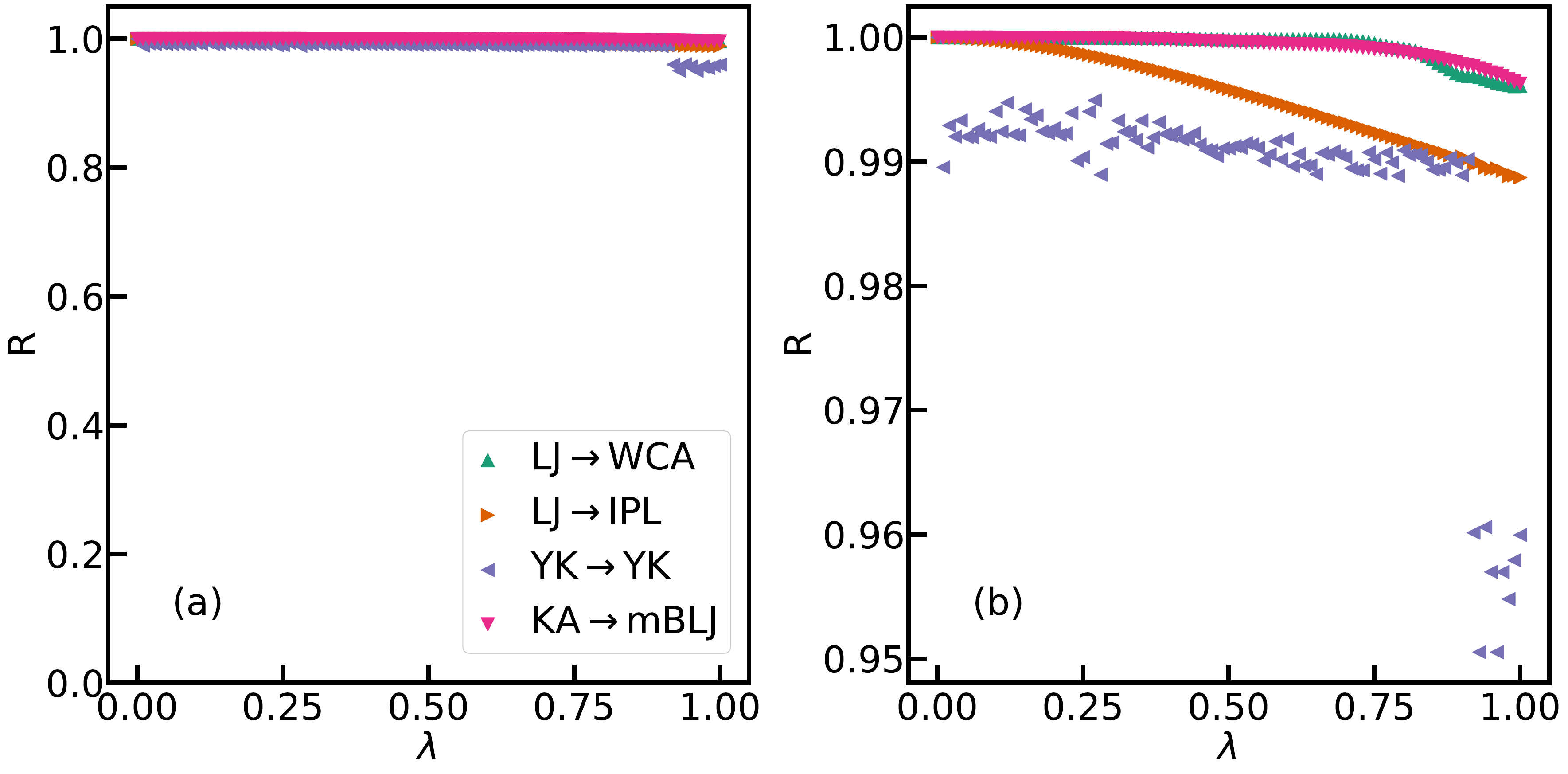}
		\caption{Correlation coefficient $R$ between the potential energies as a function of $\lambda$. Each point represents the Pearson correlation coefficient obtained by considering as independent variable the potential energies of the 32 configurations at $\lambda=0$ and as dependent variable the potential energies of the same 32 configurations evaluated at the $\lambda$ values indicated. In all cases the correlations are very strong; $R$ is mostly above 0.99 and always above 0.95. (b) shows the same data as (a).}
		\label{fig11}
	\end{figure}
	
	This paper has studied gradual interpolations between some of the most common pair-potential systems. Interpolation between two pair-potential systems was achieved by introducing a parameter $\lambda$ such that system $0$ corresponds to $\lambda=0$ and system $1$ to $\lambda=1$. The following criterion was used for checking whether two systems have the same, or almost the same, constant-potential-energy hypersurface $\Omega$: First one selects a number (here 32) of statistically independent equilibrium configurations of system $0$ at the reference state point. If the potential energies of these configurations as functions of $\lambda$ show no or only few level crossings, the two systems have the same or almost the same $\Omega$. By reference to the equivalence of $NVU$ and $NVT$ dynamics, this implies the same or almost the same structure and dynamics. The excess entropy $\Sex$ is also the same or almost the same for the two systems, while other thermodynamic quantities like the potential energy and the Helmholtz and Gibbs free energies are not expected to be invariant when interpolating between system $0$ and system $1$. We note, incidentally, that our findings confirm the quasiuniversality of simple liquids \cite{ros99,you03,you05,hey07,ram11,sch11,han13,lop13,dyr16} by the fact that all systems have almost identical $\Omega$, structure, and dynamics (compare Figs. 3-5 and 8). 
	
	We presented data for three interpolations between single-component pair-potential systems and data for an interpolation between two different versions of the KA binary LJ liquid. In all four cases there are only few ``level crossings'', which means that \eq{eq:crit_eq2} applies to a good approximation. This claim is a bit hand-waving, however. Can one can quantify to which degree \eq{eq:crit_eq2} is fulfilled? Inspired by Ref. \onlinecite{lan25} we use the Pearson correlation coefficient $R$ between the 32 different potential energies at the starting point $\lambda=0$ and at an arbitrary $\lambda\le 1$. \Fig{fig11} shows such data plotted with the ordinate axis going to zero in order to visualize how close the correlation coefficients are to unity. Most data have $R>0.99$, but the $\YtoY$ interpolation has $0.95<R<0.96$. In summary, the intuitive criterion of ``few'' level crossings translates into correlation coefficients that are indeed close to unity.
	
	An important point is that if two systems have the same $\Omega$ at two state points of same density, this does not imply the two systems have the same temperature. A crucial ingredient of the interpolation method is therefore the introduction of iterative reduced-force matching based on \eq{eq:red_force_mat}, which identifies the temperature at which the $\lambda$ system is predicted to have the same physics as system $0$ at the reference state point. 
	
	We checked for ``same physics'' by evaluating the RDF and the MSD. An obvious question is whether more complex liquid characteristics like higher-order structural measures or collective dynamic quantities like the frequency-dependent viscosity are also expected to be virtually invariant if there are only few level crossings. We have not investigated this but do expect it to be the case because of the equivalence of $NVU$ and $NVT$ dynamics \cite{NVU_I}. Another point it would be useful to look into in future works is the effect of changing the ratio of the A and B particle masses of the binary LJ system. Following the original Kob-Andersen paper \cite{kob95} we assumed identical masses, but changing the ratio will modify both the single-particle and the collective dynamics. At the same time, the above \textit{NVU}-based arguments would not be affected because the statistical Boltzmann probability are not influenced by the particle masses. Our prediction is that one would still see very similar dynamics of the shifted-potential and shifted-force versions of the system, however. We base this expectation on the fact that standard \textit{NVU} geodesic dynamics is modified for systems of different particle masses because the metric defining geodesics itself involves the particle masses, a point first discussed by Hertz \cite{NVU_III,lutzen}.
	
	The interpolation approach introduced above is general. It is not limited to pair-potential systems or to systems of same density, although these two restrictions apply to the cases studied. For future work, in order to validate our ``absence-of-level-crossing'' criterion ensuring very similar structure and dynamics after temperature adjustment, it will be important to study systems of varying density \cite{IV,dyr18a}, as well as to interpolate between non-pair potential systems, between atomic and molecular models, and between different molecular models.

	\section*{Data availability}
	The repository \href{https://zenodo.org/records/15113178}{https://zenodo.org/records/15113178} contains the data discussed in this work and the Jupyter notebook used for producing the figures.

	\begin{acknowledgments}
		The proof of concept for this work was established in a Roskilde University bachelor project by Ana Maria B. Brea, Andreas C. Martine, Claudia X. Romero, Jone E. Steinhoff,  Francisco M. F. A. S. da Fonseca, Maria B. T. Nielsen, and Victor E. S. Rasmussen. The project was supervised by J.C.D. and L.C. After graduating, Martine, Romero, Steinhoff, da Fonseca, and Nielsen contributed to the writing of the paper. This work was supported by the VILLUM Foundation's \textit{Matter} grant VIL16515.
	\end{acknowledgments}

	\clearpage
	\section*{Appendix}
	\setcounter{figure}{0}
	\setcounter{table}{0}
	\renewcommand{\thefigure}{A\arabic{figure}}
	
	This Appendix gives further simulation details, compares the dynamics of the shifted-potential and shifted-force KA versions with same (long) cutoff, and describes the iterative reduced-force-matching method.
	
	\subsection*{Simulation details}
	All simulations were performed using the GPU-optimized Molecular Dynamics code RUMD \cite{RUMD} using an $NVT$ integrator based on a leap-frog discretization of the Newtonian equation of motion coupled with a Nose-Hoover thermostat. Table \ref{tab:simdetails} lists simulation details including the number of steps "Short sim" and "Long sim" indicated as vertical dashed lines in Fig. A1 (see below).
	
	\vspace{0.5cm}
	\begin{table}[b]{}
		\centering    
		\caption{Simulation details.}
		\vspace{0.5cm}
		\begin{tabular}{c|c|c|c|c|c|c|c}
			Interpolation & N & Time step & Short sim & Long sim & Density & Ref. Temp. & Saving\\
			\hline
			$\LtoW$ & $4096$ & $0.002$ & $2^{21}$ & $2^{25}$ & $1.0$ & $2.00$ & $128$\\
			$\LtoI$ & $4096$ & $0.002$ & $2^{21}$ & $2^{25}$ & $1.0$ & $2.00$ & $128$\\
			$\YtoY$ & $4096$ & $0.010$ & $2^{20}$ & $2^{24}$ & $0.5$ & $0.02$ & $128$\\      
			%  $\LtoY$ & $4096$ & $0.001$ & $2^{24}$ & $2^{28}$ & $1.0$ & $2.00$ & $128$\\
			$\KtoL$ & $4000$ & $0.005$ & $2^{24}$ & $2^{28}$ & $1.2$ & $0.48$ & $64$ \\
		\end{tabular}
		\label{tab:simdetails}
	\end{table}
	\vspace{0.5cm}

	\begin{figure}[H]
		\centering
		\includegraphics[width=0.5\linewidth]{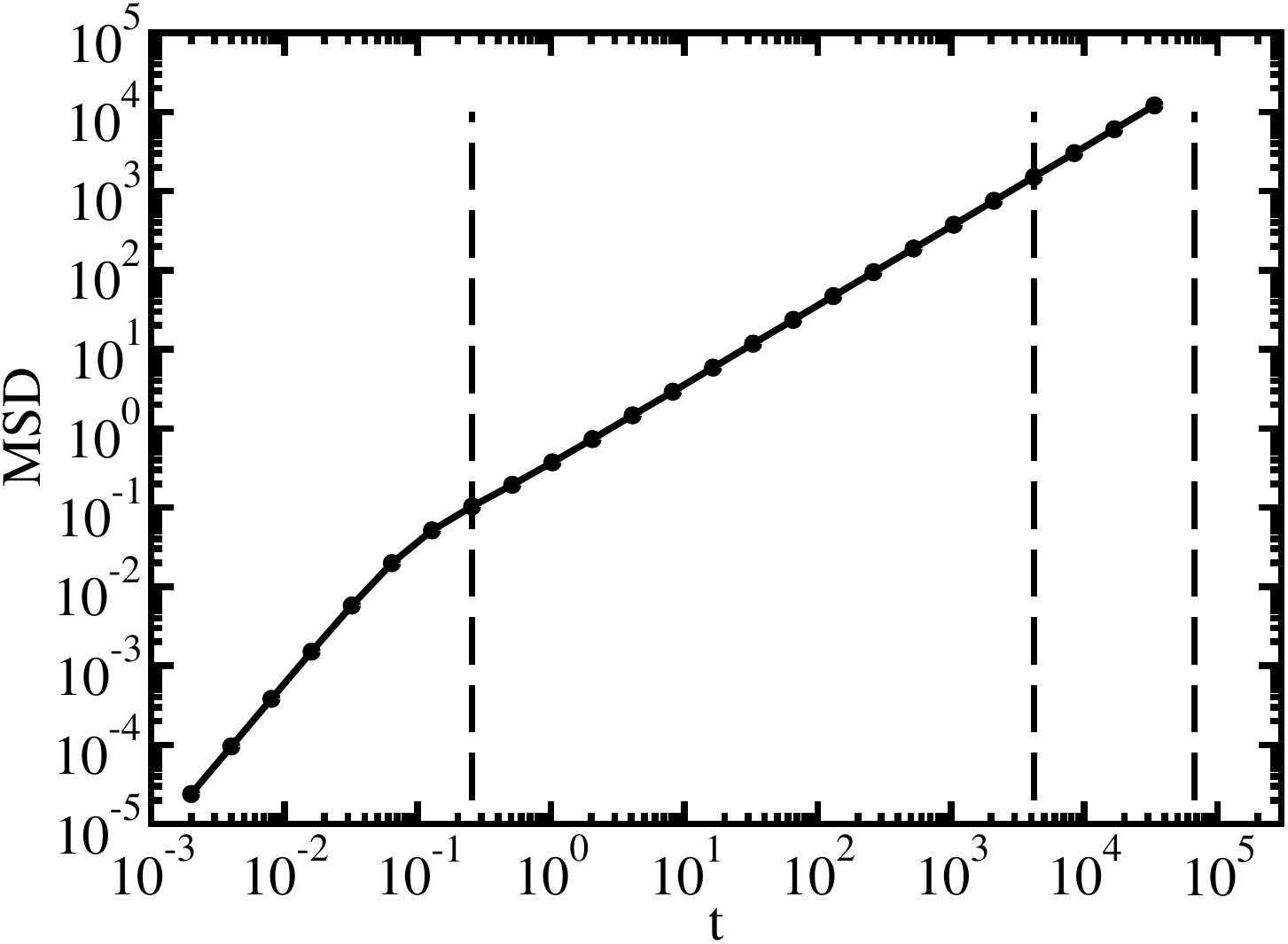}
		\caption{Time scales associated with the number of steps reported in Table 1. From left to right the vertical dashed lines correspond to how frequently the average value of the force is saved ("Saving" in Table 1) and the length of the short and the long simulations, respectively.}
		\label{figA1}
	\end{figure}

	\subsection*{Comparing the dynamics of KA systems with different cutoffs}
	\begin{figure}[H]
		\centering
		\includegraphics[width=0.5\linewidth]{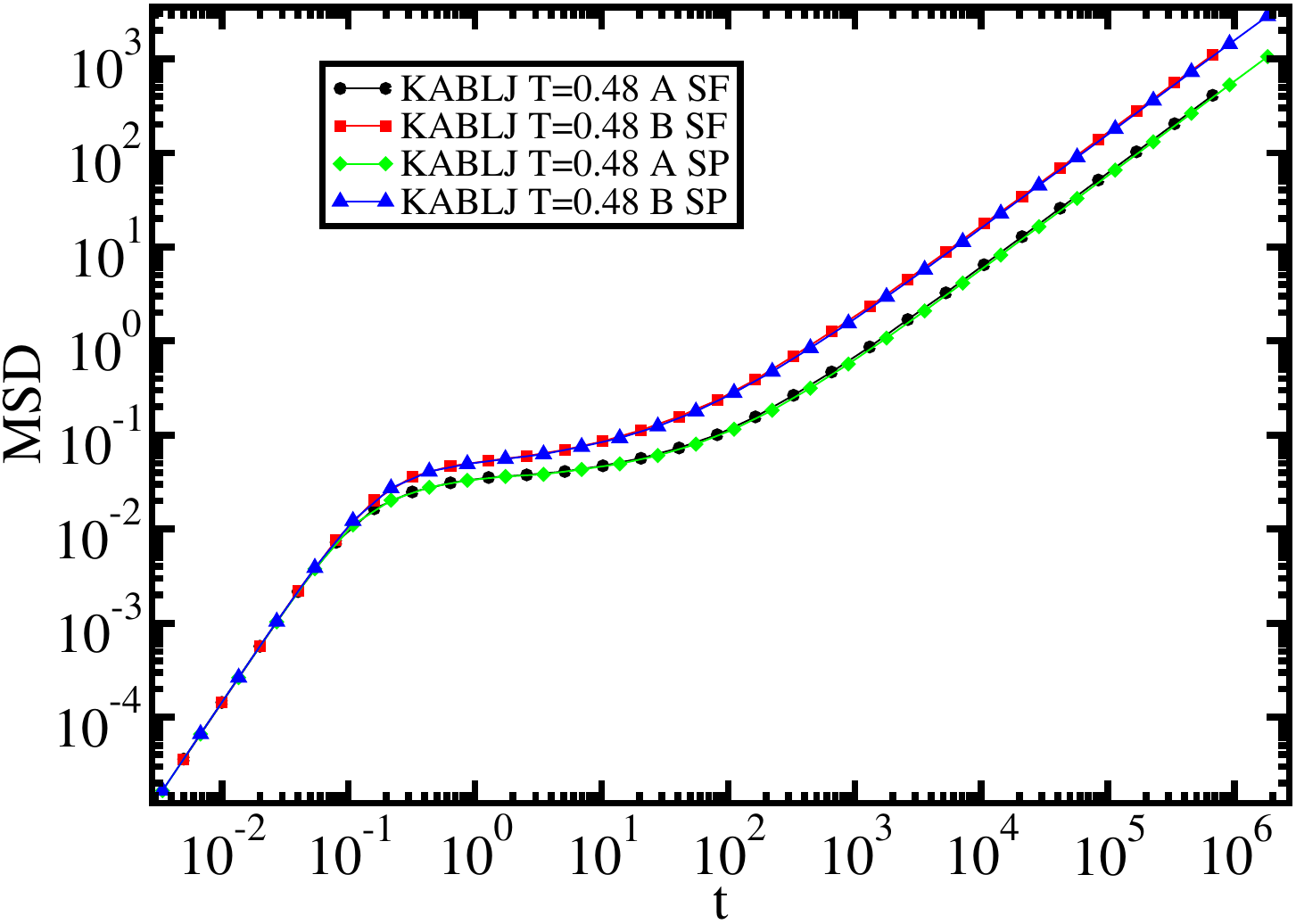}
		\caption{Comparing the A and B particle MSD of the standard KA systems with shifted-potential (SP) and shifted-force (SF) cutoffs at the state point $(\rho, T) = (1.2, 0.48)$.}
		\label{figA2}
	\end{figure}
	Figure \ref{figA2} compares the MSD of the KA system when the interaction potential is truncated by either shifted-force (SF) or shifted-potential (SP) cutoffs at the same (standard) distances. Both the A and B particle MSDs are shown. There is no visible difference between the two cutoff implementations.

	\subsection*{The reduced-force-matching procedure}\label{sec:iteraction}
	
	The reduced-force-matched temperatures (\fig{fig6}) were calculated by matching the value of the reduced force according to \eq{eq:red_force_mat} \cite{sch22}. The first step is to equilibrate the $\lambda=0$ system and simulate it for the number of steps indicated as "Long sim" in Table \ref{tab:simdetails}. From this the value of the length of the $3N$-dimensional force vector $|\mathbf{F}(\mathbf{R})|$ is sampled with the frequency indicated by "Saving". The time average of the length of the reduced force vector, ${|\mathbf{F}(\mathbf{R})|}/{k_B T}$, is evaluated and used in the following steps as reference (note that the reduced force vector in general includes a density factor \cite{IV}, which for simplicity is omitted in this paper because density is kept constant).
	
	The temperature corresponding to the same reduced-force vector at a new value of $\lambda$ is found using the following iterative procedure.
	
	\begin{enumerate}
		\item The length of the reduced force vector is evaluated for the new value of $\lambda$.
		\item A new temperature, $T(\lambda)$, is found from ${|\mathbf{F}(\mathbf{R})|_{\lambda}}/{k_B T(\lambda)} = {|\mathbf{F}(\mathbf{R})|_{\lambda=0}}/{k_B T(\lambda=0)}$.
		\item A new simulation of length "Short sim" is run at $T(\lambda)$, after which the length of the reduced-force vector is evaluated and compared to the reference value ${|\mathbf{F}(\mathbf{R})|_{\lambda=0}}/{k_B T(\lambda=0)}$
		\item Steps 2 and 3 are repeated until the relative difference between the lengths of the reduced force vector is below $5\cdot 10^{-6}$.
	\end{enumerate}
	
	The need for this iterative procedure arises from the fact that the theory is only approximate. For selected values of $\lambda$ ($0.2$, $0.4$, $0.6$, $0.8$, $1.0$) a longer simulation (of "Long sim" steps) is run to obtain better statistic on the RDF and the MSD; these data are the ones shown in the figures of the main paper. The $\lambda$ parameter varies in all cases between $0$ and $1$, an interval that is divided into steps of $0.05$. This is why there are $21$ points for $T(\lambda)$ in Fig. \ref{fig6} (short simulations) and only 6 for $\tilde{D}(\lambda)$ (long simulations).
	
	\newpage

\end{document}